\documentclass[prx,twocolumn,groupeaddress]{revtex4-1}

\usepackage{amsmath}
\usepackage{amssymb}
\usepackage{xspace}
\usepackage{graphicx}
\usepackage{grffile}
\usepackage{nicefrac}
\usepackage{array}
\usepackage{color}

\graphicspath{{figs/}}
\newcolumntype{L}[1]{>{\raggedright\let\newline\\\arraybackslash\hspace{0pt}}m{#1}}
\newcolumntype{C}[1]{>{\centering\let\newline\\\arraybackslash\hspace{0pt}}m{#1}}
\newcolumntype{R}[1]{>{\raggedleft\let\newline\\\arraybackslash\hspace{0pt}}m{#1}}
\newcommand{\bw}[1]{\raisebox{1.5ex}[-1.5ex]{#1}}

\begin{document}

\title{Multiorbital processes rule the 
Nd$_{1-x}$Sr$_x$NiO$_2$ normal state} 
\author{Frank Lechermann}
\affiliation{I. Institut f{\"u}r Theoretische Physik, Universit{\"a}t Hamburg, 
Jungiusstr. 9, 20355 Hamburg, Germany}

\begin{abstract}
The predominant Ni-multiorbital nature of infinite-layer neodynium nickelate
at stoichiometry and with doping is revealed. We investigate the
correlated electronic structure of NdNiO$_2$ at lower temperatures and show
that first-principles many-body theory may account for Kondo(-lattice) 
features. Yet those are not only based on localized Ni-$d_{x^2-y^2}$ and a 
Nd-dominated self-doping band, but heavily builds on the participation of 
Ni-$d_{z^2}$ in a Hund-assisted manner. In a tailored three-orbital study, 
the half-filled regime of the former inplane Ni orbital remains 
surprisingly robust even for substantial hole doping $\delta$. Reconstructions of 
the interacting Fermi surface designate the superconducting region within the 
experimental phase diagram. They furthermore provide clues to recent Hall 
measurements as well as to the astounding weakly-insulating behavior at larger 
experimental $\delta$.
Finally, a strong asymmetry between electron and hole doping, with a revival
of Ni single-orbital features in the former case, is predicted. Unlike cuprates, 
superconductivity in Nd$_{1-x}$Sr$_x$NiO$_2$ is of distinct multiorbital kind, 
building up on nearly localized Ni-$d_{x^2-y^2}$ and itinerant Ni-$d_{z^2}$.
\end{abstract}

\maketitle

\section{Introduction} 
The discovery of superconductivity in Sr-doped thin films of infinite-layer (IL)
NdNiO$_2$ with a $T_{\rm c}$ in the 10\,K range marks a further milestone in the 
investigation of transition-metal oxides~\cite{li19}. Recent continuing
experimental works provided details on the thin-film growth on a SrTiO$_3$
substrate, and moreover yield a phase diagam with doping~\cite{li20,lee20,zen20}.
There, the superconducting region is placed in the range 
$0.125\lesssim x\lesssim0.25$ within the Nd$_{1-x}$Sr$_x$NiO$_2$ system. 
Notably, that area is neighbored by weakly-insulating regions on either side of 
the doping range. Nonsurprisingly, the original findings~\cite{li19} are already 
covered by many theoretical 
works~\cite{gu19,hep20,hu19,zha20-2,zhav20,wer20,lec20,zhou19,hir19,liu19,jia19,jiang19,rye19,wu19,nom19,bot20,zha20,si20,kit20,cho20,kar20,ole20,leo20,bee20,nic20},
with also important earlier studies~\cite{ani99,lee04,cha08,han09} on similar 
nickelate systems. 

\begin{figure}[t]
\includegraphics*[width=8.5cm]{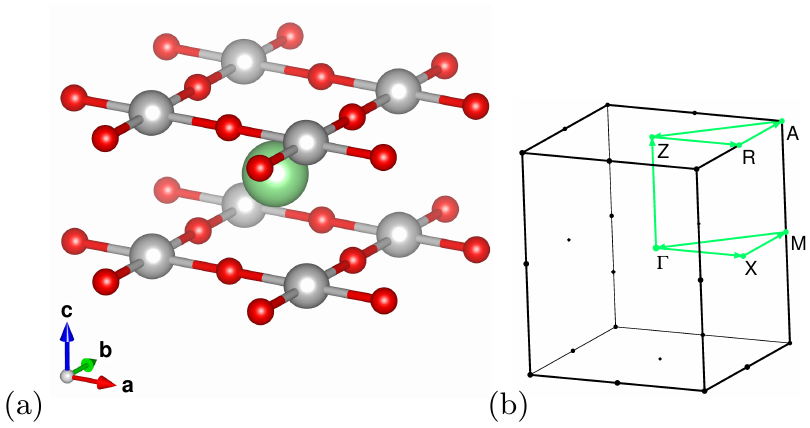}
\caption{(color online) Infinite-layer NdNiO$_2$. (a) Crystal structure with Nd 
(large green), Ni (blue) and O (small red) sites. (b) Brillouin zone with 
high-symmetry points in $k_z=0$ and $k_z=1/2$ plane. Green lines depict the path
for plots of the spectral function $A({\bf k},\omega)$.}
\label{fig:struc}
\end{figure}

Up to now, three (interlinked) key questions are associated with the challenging 
physics of superconducting nickelates. The first one deals with the basic comparison to 
high-$T_{\rm c}$ layered cuprates. Though seemingly akin, IL nickelate (see 
Fig.~\ref{fig:struc}a) with the unusual Ni$^+$ formal oxdiation state shows 
prominent differences at stoichiometry: NdNiO$_2$ is weakly metallic and does not 
exhibit magnetic ordering down to the lowest measured temperatures~\cite{li19}.
Theoretical calculations furthermore show that it has an additional self-doping (SD) 
band crossing the Fermi level and that the charge-transfer character is weaker 
than in cuprates~\cite{jia19,lec20}. Second, strong correlations within the 
Ni-$d_{x^2-y^2}$ state together with an existing SD band, raises the question about 
Kondo(-lattice) physics at low temperature. An experimentally found~\cite{li19} 
resistivity upturn below about $T\sim 70$\,K might indeed be indicative for 
related processes. The third, highly debated, question is most relevant for the 
superconducting mechanism and deals with the issue of deciding
low-energy physics based on single-Ni-orbital processes of Ni-$d_{x^2-y^2}$ 
kind~\cite{wu19,nom19,bot20,zha20,si20,kit20} versus processes of
multi-Ni-orbital kind~\cite{lee04,hu19,zhav20,wer20,lec20} at stoichiometry and
with finite hole doping from replacing Nd by Sr.  

In a previous work~\cite{lec20}, we focussed on the one-to-one comparison of IL nickelates
to structural-akin cuprates. The present study uncovers crucial Ni-multiorbital mechanisms 
for Kondo physics at stoichiometry and for the accentuation of the superconducting region 
with hole doping. 

First-principles many-body theory and a realistic three-band Hamiltonian investigation 
are utilized to establish our current understanding of IL nickelates. Besides the 
significance of Ni-$d_{x^2-y^2}$ and the Nd-dominated SD band, we unveil the decisive 
role of Ni-$d_{z^2}$, both for a Kondo(-lattice) behavior at stoichiometry and for the
characterization of the normal-state correlated electronic structure with hole doping 
$\delta$. While the
Ni-$d_{x^2-y^2}$ Fermi-surface sheet in the $k_z=0$ plane does hardly evolve with 
increasing $\delta$, an additional Ni-$d_{z^2}$ sheet becomes available in the
$k_z=1/2$ plane for dopings in the region of the onset of superconductivity. 
On the other hand, further changes of the Fermi surface at larger hole doping may be 
connected to the experimentally found weakly-insulating behavior. The given 
Fermi-surface reconstructions with doping are furthermore in line with the 
experimentally established Hall-data based multiband picture of an evolution 
from electron-like to hole-like transport. In a final step, we compare hole doping
with theoretical electron doping and detect a stronger single-orbital character
of Ni-$d_{x^2-y^2}$ flavor on the electron-doped side.

\section{Methods}
In essence, three methodologies are put into practise in this work. First,
the charge self-consistent combination~\cite{gri12} of density functional 
theory (DFT), self-interaction correction (SIC) and dynamical mean-field theory 
(DMFT), i.e., the so-called DFT+sicDMFT framework~\cite{lec19}, is used to provide
a realistic approach to the temperature-dependent correlated electronic
structure of NdNiO$_2$. The complete Ni$(3d)$ shell enters the correlated
subspace of the DMFT impurity problem. Coulomb interactions on oxygen are  
described within SIC, and are incorporated in the O pseudopotential~\cite{kor10}. 
A mixed-basis pseudopotential code~\cite{els90,lec02,mbpp_code} takes care of the
DFT part in the local density approximation (LDA). The SIC is applied to the O$(2s)$ 
and the O$(2p)$ orbitals via weight factors $w_p$ (see Ref.~\onlinecite{kor10} 
for more details). While the O$(2s)$ orbital is by default fully corrected with $w_p=1.0$, 
the common choice~\cite{kor10,lec19} $w_p=0.8$ is used for O$(2p)$ orbitals. 
Hence the oxygen states are treated beyond conventional DFT+DMFT, but still not on
full eye level with Ni$(3d)$. Cluster calculations for transition-metal oxides
(e.g. Refs.~\onlinecite{fuj84,elp92,tag08}) have been put forward in this direction, 
but those schemes then often suffer from other problems, such as e.g. parametrization 
issues and breaking of translational invariance.

The Nd$(4f)$ states are put in the pseudopotential frozen core, since they appear
irrelevant for the key physics of superconducting IL nickelates~\cite{zha20}. 
Continuous-time quantum Monte Carlo in hybridzation expansion~\cite{wer06} as 
implemented in the TRIQS code~\cite{par15,set16} is utilized to address the 
DMFT problem. The DMFT correlated subspace is governed by a five-orbital full 
Slater Hamiltonian applied to the Ni projected-local orbitals~\cite{ama08}. 
The projection is performed on the $6+5+1=12$ Kohn-Sham states above the dominant 
O$(2s)$ bands, associated with O$(2p)$, Ni$(3d)$ and the self-doping band.
A Hubbard $U=10$\,eV and a Hund's exchange $J_{\rm H}=1$\,eV prove reasonable for 
this large-energy window treatment of the given late transition-metal oxide~\cite{lec20}. 
The fully-localized-limit double-counting scheme~\cite{ani93} is applied. 
The DFT+sicDMFT calculations are performed in the 
paramagnetic regime. Maximum-entropy and Pad{\'e} methods are employed for the 
analytical continuation from Matsubara space onto the real-frequency axis. 
Stoichiometric lattice parameters are overtaken from experiment~\cite{li19}.

Second, we employ the maximally-localized Wannier-function (MLWF) 
formalism~\cite{mar97,sou01} to construct an effective three-orbital low-energy 
Hamiltonian for NdNiO$_2$, to be utilized at stoichiometry and with finite doping. 
Details of the construction are provided in section~\ref{sec:mod}.

Third, the derived Hamiltonian is solved by the rotationally invariant
slave-boson (RISB) scheme~\cite{li89,lec07,isi09,lan17,pie18,fac18} on the mean-field 
level. The RISB electronic self-energy is local (or extendable via cluster techniques)
and consists of a term linear in frequency as well as a static part. It thus lacks the 
full frequency dependence of the DMFT self-energy, but is still well suited (here at 
formal $T=0$) for a large class of correlated materials problems. For details on the 
computation of quantities such as the quasiparticle weight $Z$ or 
local spin correlations see e.g. Refs.~\onlinecite{lec07,pie18}.

\section{Kondo signature in N\MakeLowercase{d}N\MakeLowercase{i}O$_2$ from  
comprehensive realistic DMFT\label{sec:kondo}}
\begin{figure}[b]
\includegraphics*[width=8.5cm]{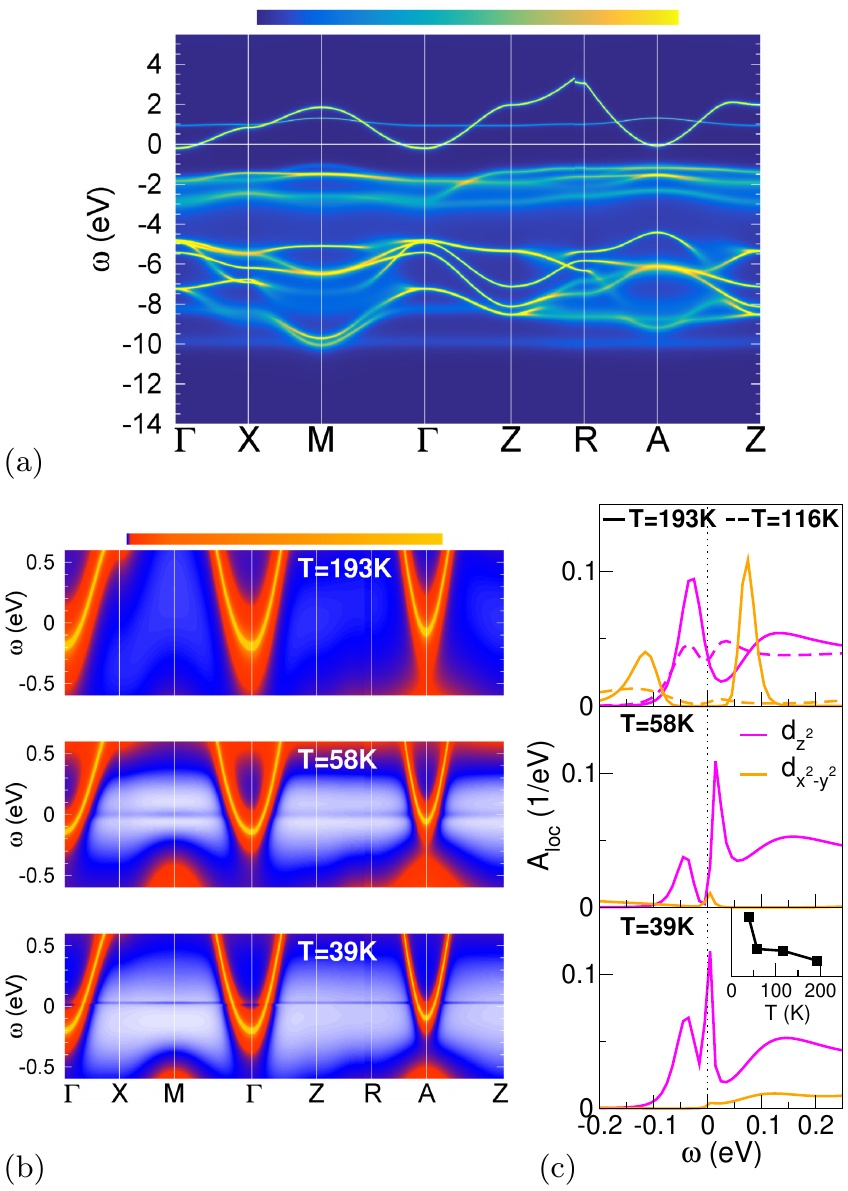}
\caption{(color online) Evolution of the 
DFT+sicDMFT spectral data for NdNiO$_2$ with lowering temperature.
(a) {\bf k}-resolved spectrum $A({\bf k},\omega)$ along high-symmetry lines in the 
Brillouin zone over a wide energy range at $T=193$\,K (from Ref.~\onlinecite{lec20}).
(b) Low-energy $A({\bf k},\omega)$ for three different temperatures.
(c) $T$-dependent  Ni-$d_{z^2}$ and Ni-$d_{x^2-y^2}$ local spectral function $A_{\rm loc}$ 
close to the Fermi level. The small inset in the bottom part shows the two-orbital
integrated spectral weight in the energy window $[-0.05,0.05]$\,eV.}
\label{fig:spec}
\end{figure}
In Ref.~\onlinecite{lec20} we studied the interacting electronic
structure of NdNiO$_2$ using the DFT+sicDMFT method at the system temperature $T=193$\,K.
At the reasonable large interaction strength
$U=10$\,eV, the half-filled Ni-$d_{x^2-y^2}$ band is Mott insulating and does not 
participate in the Fermi surface. This is shown in Fig.~\ref{fig:spec}a, where only the 
electron pockets of the self-doping band cross the Fermi level 
$\varepsilon^{\hfill}_{\rm F}$ in the $k_z=0$ plane around $\Gamma$ and in the
$k_z=1/2$ plane around A. The SD band is of mixed Ni-Nd character, namely Nd-$d_{z^2,xy}$
and especially relevant Ni-$d_{z^2}$ contribution to the $\Gamma$ 
pocket~\cite{nom19,lec20}. A more detailed discussion of the orbital hybridizations can
be found in Ref.~\onlinecite{lec20}.

At lower temperatures, a coupling of the localized Ni-$d_{x^2-y^2}$-based spin-1/2 and 
the remaining itinerant degrees of freedom is suggested from the strong-coupling 
situation. An underscreening scenario is then expected because of the low filling of 
the electron pockets. But importantly, there is zero nearest-neighbor hopping between 
Ni-$d_{x^2-y^2}$ and Nd-$d_{z^2,xy}$ (e.g. Ref.~\onlinecite{lec20}). Thus an 
intriguing Kondo(-lattice) picture is expected. Note however that it is computationally 
challenging to reach very low temperatures within our five-Ni-orbital realistic 
DMFT framework.
\begin{figure}[t]
\includegraphics*[width=8.5cm]{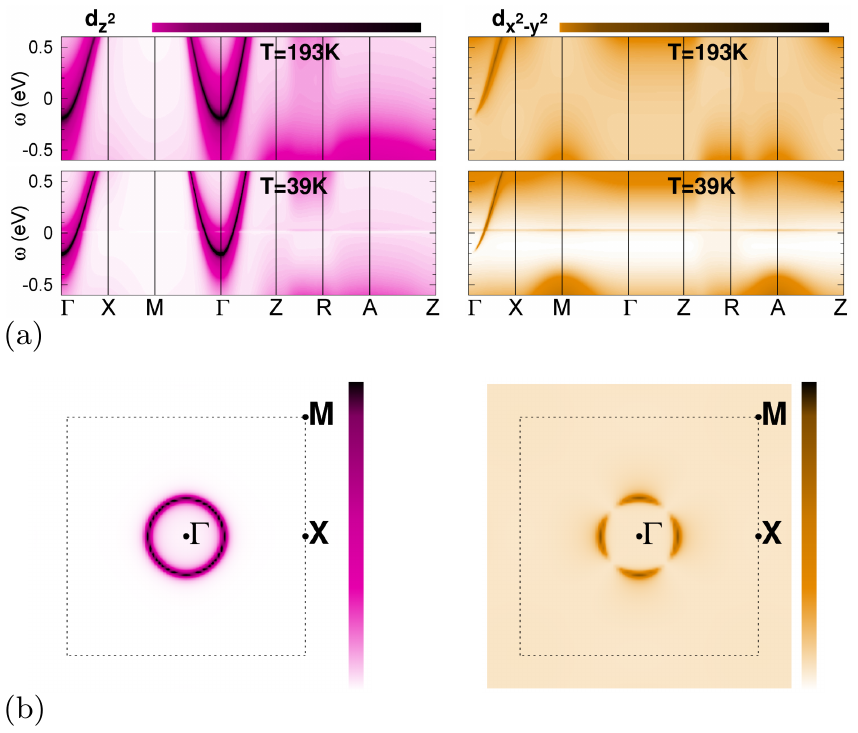}
\caption{(color online) Orbitals weights (i.e. 'fatband' picture) for Ni-$d_{z^2}$ (left)
and Ni-$d_{x^2-y^2}$ (right) character in the {\bf k}-resolved
DFT+sicDMFT spectrum of NdNiO$_2$.
(a) Along high-symmetry lines in the Brillouin zone at $T=193$\,K and $T=39$\,K.
(b) On the Fermi surface in the $k_z=0$ plane at $T=193$\,K. Note that the Ni-$d_{z^2}$
colorscale maximum is 30 times larger than the Ni-$d_{x^2-y^2}$ one.}
\label{fig:fat}
\end{figure}

Still, Fig.~\ref{fig:spec}b displays the evolution of the {\bf k}-resolved correlated 
electronic structure at low energy with decreasing $T$. At $T=58$\,K, 
a flat-dispersion feature appears in the spectral function $A({\bf k},\omega)$ 
close to $\varepsilon^{\hfill}_{\rm F}$, which further settles below 40\,K. The location of 
the electron pockets of the SD band remains rather $T$-independent. It is very tempting to
link this lowest-energy crossing of a flat-band feature and itinerant, seemingly weakly 
correlated dispersions, to a Kondo scenario. Yet the spectral intensity of the flat feature 
remains very weak at these reachable temperatures. The orbital-resolved local
spectrum depicted in Fig.~\ref{fig:spec}c renders obvious, that the main Ni-derived
spectral weight in this region is of Ni-$d_{z^2}$ character. At $T=193$\,K, a single
Ni-$d_{z^2}$ quasiparticle peak is located just below $\varepsilon^{\hfill}_{\rm F}$
at $\sim -30$\,meV. Lowering temperature first leads to a second peak above the Fermi 
level that gradually becomes sharper and shifts towards $\varepsilon^{\hfill}_{\rm F}$.
Hence the 'Kondo resonance' property, at least in that $T$ range, is carried by the
Ni-$d_{z^2}$ character. This points to a relevant participation of Ni-$d_{z^2}$ in
the given low-energy physics. The local spectral weight of Ni-$e_g$ kind 
near $\varepsilon^{\hfill}_{\rm F}$ strongly grows below 
$T=58$\,K (see inset at bottom in Fig.~\ref{fig:spec}c), in good agreement with the 
experimental temperature scale for the onset of the resistivity increase. 
The orbital-resolved electron count is hardly dependent on $T$ and reads 
$\{n_{z^2}, n_{x^2-y^2}\}=\{1.845,1.065\}$ for Ni.

To investigate the roles of the two Ni-$e_g$ orbitals further, we also extracted their
{\bf k}-resolved orbital characters (i.e. 'fatbands') in the interacting regime, as shown
in Fig.~\ref{fig:fat}a. It is first once more confirmed that there is no Ni-$e_g$
contribution to the electron pocket around A. Second, the Ni-$d_{x^2-y^2}$ orbital weight
at low-energy is way smaller than the Ni-$d_{z^2}$ one. The Ni-$d_{x^2-y^2}$ origin of the
flat feature at low $T$ is still obvious. Third, the Ni-$d_{x^2-y^2}$ orbital displays a 
low-energy hybridization with Ni-$d_{z^2}$ within the $\Gamma$ pocket, however interestingly, 
only along $\Gamma-$X in the chosen Brillouin-zone path. This specific direction-dependent 
hybridization is clearly visible already at $T=193\,$K and by close inspection even observable 
on the LDA level. Its structure is yet better revealed when plotting the Ni-$d_{x^2-y^2}$ 
orbital weight along the $k_z=0$ Fermi surface (see Fig.~\ref{fig:fat}b). The corresponding 
weight is strictly zero 
\begin{figure}[t]
\includegraphics*[width=8.5cm]{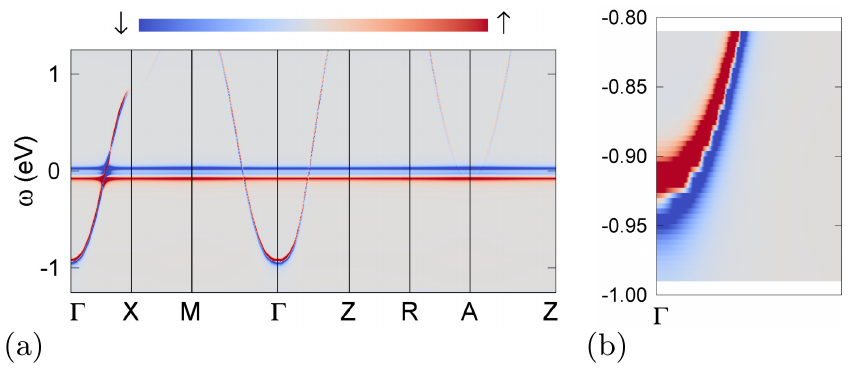}
\caption{(color online) Spin contrast 
$A_{\uparrow}({\bf k},\omega)-A_{\downarrow}({\bf k},\omega)$ 
from a one-step DFT+sicDMFT calculation with spin-polarized Ni self-energies (see text).
(a) Energy window around Fermi level. (b) Blow up close to $\Gamma$, to extract the
spin splitting of the electron pocket.}
\label{fig:kondo}
\end{figure}
along $\Gamma-$X, and along $\Gamma-$Z from Fig.~\ref{fig:fat}a. Since the oxygen positions
in real space correspond to the $\Gamma-$X direction, we attribute this Ni-Ni hybridization
to explicit hopping via oxygen. It appears relevant for the present Kondo
scenario since it provides the sole explicit Ni-$d_{z^2}$ connection between the single
level and the $\Gamma$ pocket at lower temperature (cf. Fig.~\ref{fig:fat}a).
As an inspection below $T=58$\,K, a Fermi-surface maximum on the circular sheet appears
along $\Gamma-$X and splits perpendicular to the $\Gamma-$X direction, respectively 
(not shown).

In order to eventually also connect to the spin degree of freedom in a simplest manner, we 
choose a linear-response(-kind) calculation. For $T=39$\,K, we strongly spin split the 
paramagnetic Ni-$d_{x^2-y^2}$ occupation by hand and perform a single DFT+sicDMFT step
allowing for spin-polarized Ni self-energies. This one-step 'magnetic-order' calculation 
leads to a spin contrast in the low-energy spectral function as shown in 
Fig.~\ref{fig:kondo}a. As expected, the single level, mirroring the localized-spin 
behavior, is split into a larger-occupied spin-up branch and a smaller-occupied
spin-down part (due to the performed spin splitting by hand). 
Important is however the affect of this
splitting onto the itinerant pockets, since it may deliver information about the Kondo(-like)
exchange coupling. Also here, the A pocket shows only weak spin signature. Yet the $\Gamma$
pocket displays sizable spin splitting, hence is 'Kondo' affected by the localized
Ni-$d_{x^2-y^2}$ spin. As one can easily see, that pocket's spin splitting is the other way
around, i.e., the spin-up band is shifted to somewhat higher energies. This indeed points to
an antiferromagnetic (AFM) exchange coupling between Ni-$d_{x^2-y^2}$ and the SD band.

The size of the SD spin splitting at $\Gamma$ amounts to 40\,meV (see Fig.~\ref{fig:kondo}b), 
which translates in a Kondo coupling $J_{\rm K}\sim 120$\,meV. This is a quite large value 
for $J_{\rm K}$, but indeed in agreement with recent estimates~\cite{zha20,yan08}. Notably, 
a Ni-$d_{z^2}$ supported AFM exchange is not in contradiction with Hund's first rule.
In fact in the regime of highly occupied Ni-$d_{z^2}$, a {\sl local} $S=1$ triplet from 
$d_{z^2}$ and $d_{x^2-y^2}$ favors an additional {\sl itinerant} $d_{z^2}$ which is
AFM aligned to $d_{x^2-y^2}$. As a final observation, the spin-contrast signal is 
highlighted at the $\Gamma-$X crossing of the single level and the $\Gamma$ pocket, 
underlining once more the relevant Ni-$d_{z^2}$ role.

\section{Low-energy study of the correlated electronic structure}
To proceed on the electronic states in IL nickelates in a more general and flexible way, 
let us turn to an explicit low-energy description, where we focus on a 
minimal set of degrees of freedom that are essential for the key electronic processes 
at lower temperature. For instance, the nearly completely-filled Ni-$t_{2g}$ states are 
fully included in the DFT+sicDMFT study, yet in a first step, their effect onto the 
essential physics at low energy may be cast into a formulation based on integrating out  
those orbital degrees of freedom. 
\begin{figure}[t]
\includegraphics*[width=8.5cm]{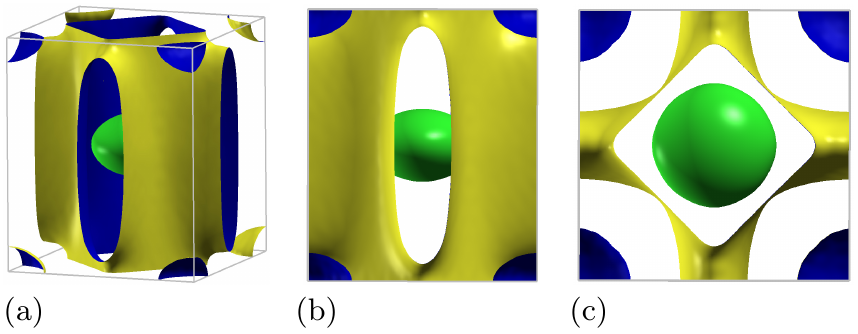}
\caption{(color online) LDA Fermi surface of pristine NdNiO$_2$. (a) 3D view, 
(b) view along $k_x$ and (c) view along $k_z$.}
\label{fig:fermi}
\end{figure}

\subsection{Minimal Hamiltonian\label{sec:mod}}
For the selection of the truly relevant degrees of freedom in an unbiased manner, it makes 
sense to be guided by the previous more comprehensive DFT+sicDMFT picture. From the results
of Ref.~\onlinecite{lec20} and those from the previous section~\ref{sec:kondo}, the natural
conclusion has to be that we definitely need Ni-$d_{z^2}$, Ni-$d_{x^2-y^2}$ and the SD band. 
The SD band carries important weight from Nd-$d_{z^2,xy}$, which then would amount to a 
four-orbital Hamiltonian. Thats not unfeasable, but one wonders if that is truly 'minimal', 
also in a model sense for general IL nickelates.
For instance, we learned that the A pocket, where the Nd-orbital contribution is most 
substantial~\cite{lec20}, is apparently not as important as the $\Gamma$ pocket in the 
low-temperature interacting regime. It may therefore be sufficient for a truly minimal
setting to merge the Nd$(5d)$ part at low energy with the remaining Ni-$t_{2g}$, Ni($4s$) 
and O$(2p)$ contributions and ally them altogether in a 'stand-alone bath' degree of 
freedom coupled to the Ni-$e_g$ orbitals. 
\begin{figure}[t]
\includegraphics*[width=8.5cm]{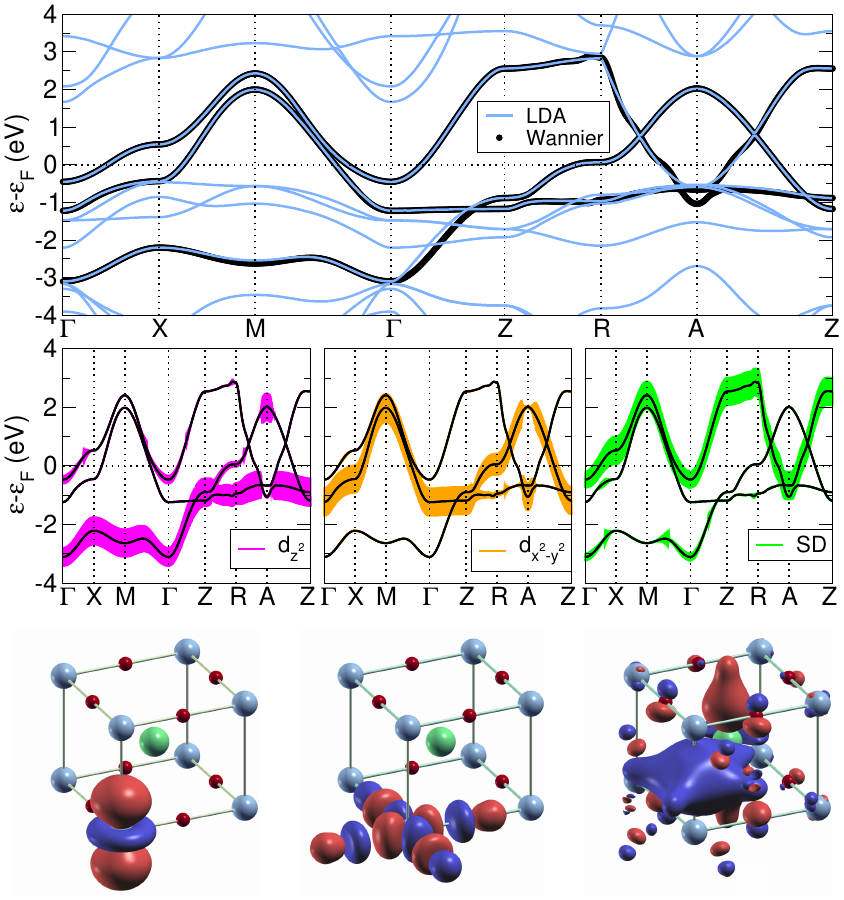}
\caption{(color online) Wannier properties of the three-band model for IL nickelate.
Top: Low-energy LDA bands of NdNiO$_2$ (lightblue) and Wannier dispersion (black)
along high-symmetry lines in the Brillouin zone.
Middle: Orbital weights on the Wannier dispersion (i.e. 'fatband' picture); from
left to right: Ni-$d_{z^2}$, Ni-$d_{x^2-y^2}$ and SD orbital.
Bottom: Constant-value surfaces of the three Wannier orbitals (aligned as in middle row).}
\label{fig:wann}
\end{figure}

We thus arrive at a three-orbital Hamiltonian, based on Ni-$d_{z^2}$, Ni-$d_{x^2-y^2}$ and 
a third orbital which we still call 'SD', but understandably now carries all 
the non-Ni-$e_g$ contributions at low energy. Concretely, the complete Hamiltonian 
reads
\begin{equation}
H_{\rm min}
= H^{\hfill}_{\rm kin} + \sum_i\left( H^{(i)}_{\rm int} + H^{(i)}_{\rm orb}\right)\;,
\label{eq:ham1}
\end{equation}
using the label $i$ for the unit cell. It incorporates the kinetic part $H_{\rm kin}$, 
the local-interacting part $H_{\rm int}$ and a further local, orbital contribution 
$H_{\rm orb}$. The kinetic Hamiltonian, here liberated from all local terms, may be 
written with hoppings $t$ as
\begin{equation}
H^{\hfill}_{\rm kin} = \sum_{i\ne j,mm',\sigma} t^{\hfill}_{ij,mm'}
\,c^\dagger_{im\sigma} c^{\hfill}_{jm'\sigma}\;,
\label{eq:hamkin}
\end{equation}
where $m,m'=\mbox{Ni-}d_{z^2},\mbox{Ni-}d_{x^2-y^2},\mbox{SD}$ and
$\sigma=\uparrow,\downarrow$.
For $H_{\rm int}$ we choose a canonical Slater-Kanamori
form to describe local interactions among the Ni-$e_g$ orbitals, i.e.
\begin{eqnarray}      
H^{(i)}_{\rm int}&=&U\sum_{m=z^2,x^2\mbox{-}y^2} n_{m\uparrow}n_{m\downarrow}    
+\sum \limits _{\sigma}              
\Big\{U' \, n_{z^2 \sigma} n_{x^2\mbox{-}y^2 \bar \sigma}\nonumber\\     
&&+\, U'' \,n_{z^2 \sigma}n_{x^2\mbox{-}y^2 \sigma}+           
J_{\rm H}\, c^\dagger_{z^2 \sigma} c^\dagger_{x^2\mbox{-}y^2 \bar\sigma} 
c^{\hfill}_{z^2 \bar \sigma} c^{\hfill}_{x^2\mbox{-}y^2 \sigma}\nonumber\\    
&&+\left. J_{\rm H}\,c^\dagger_{z^2 \sigma} c^\dagger_{z^2 \bar \sigma}  
 c^{\hfill}_{x^2\mbox{-}y^2 \bar \sigma} c^{\hfill}_{x^2\mbox{-}y^2 \sigma}\right\}\;,
\label{eq:intham}          
\end{eqnarray} 
and $U'=U-2J_{\rm H}$, $U''=U-3J_{\rm H}$.
Thus notably within the present choice, explicit interactions within the 
SD orbital, or between Ni-$e_g$ and SD, are set to zero in our minimal realistic 
modeling.
The final $H^{\rm orb}$ does not only deal with the local sum 
$\varepsilon^{\rm loc}$ of kinetic energies, but importantly also has to 
take care of double counting due to a depiction of strongly interacting Ni-$e_g$ 
coupled to a weakly interacting SD state. Furthermore, the SIC description in the
DFT+sicDMFT calculations took care of relevant interactions on O$(2p)$ in this
late transition-metal oxide, which now enters also the double-counting term.
It proves therefore favorable in this minimal context to split 
the double-counting (DC) correction into two parts:
a negative shift of the SD state through a potential $\mu_{\rm SD}$ in the 
interacting case, and a standard DC correction for explicitly interacting 
Ni-$e_g$. For the latter we choose again the fully-localized-limit 
form~\cite{ani93}. The last term in~(\ref{eq:ham1}) is hence given by
 \begin{eqnarray}
H^{(i)}_{\rm orb}&=&\sum_{mm',\sigma} \varepsilon^{^{\rm loc}}_{mm'}
\,c^\dagger_{m\sigma} c^{\hfill}_{m'\sigma}
-\mu_{\rm SD}\sum_{\sigma}n^{\hfill}_{{\rm SD}\sigma}
\nonumber\\
&&\hspace*{-2cm}-\hspace*{-0.4cm}
\sum_{m=z^2,x^2\mbox{-}y^2}\left[U\left(\tilde{n}_{e_g}-\frac{1}{2}\right)-
J_{\rm H}\left(\tilde{n}_{e_g\sigma}-\frac{1}{2}\right)\right]n_{m\sigma}\;.
\label{eq:hamloc}
\end{eqnarray}
As we will utilize the minimal Hamiltonian without charge self-consistency,
the $\tilde{n}$ occupations in~(\ref{eq:hamloc}) refer to $e_g$ fillings at 
$U=0$ (in the following, '$U=0$' is understood as $U=J_{\rm H}=0$).
The potential $\mu_{\rm SD}$ for the SD state is not perfectly 
straightforward, since therein accumulate various effects, such as electrostatics, 
interaction with O$(2p)$, etc. Since NdNiO$_2$ for the present minimal three-orbital 
Hamiltonian is located at total half filling $n=3$, the most canonical choice is 
provided by $\mu_{\rm SD}=U/2$ (e.g. Ref.~\onlinecite{schu12}). 
\begin{table}[t]
\begin{tabular}{L{3.25cm}|R{1cm} R{1.25cm} R{1cm} R{1cm} }
  orbitals                         & $\varepsilon^{\rm loc}$ & $t_{12}^{(x)}$  & $t_{12}^{(z)}$  & $t_{13}^{(x)}$\\ \hline
Ni-$d_{z^2}$, Ni-$d_{z^2}$         & -1479 &  -1   & -398 & 0\\[0.1cm]
Ni-$d_{x^2-y^2}$, Ni-$d_{x^2-y^2}$ &  232  &  -387 & -30  & -50\\[0.1cm]
SD, SD                             &  1191 &  -13  & -229 & 12 \\[0.1cm]
Ni-$d_{z^2}$, SD                   &  92   &  125  & 77   & -8 \\[0.1cm]
Ni-$d_{z^2}$, Ni-$d_{x^2-y^2}$     &  0    &  41   & 0    & 9 \\[0.1cm]
Ni-$d_{x^2-y^2}$, SD               &  0    &  17   & 0    & 13\\
\end{tabular}
\caption{Local single-particle terms and selected hoppings of the derived Wannier Hamiltonian.
All energies in meV.}
\label{tab:hopp}
\end{table}

For a concrete representation of the described Hamiltonian form, maximally-localized 
Wannier functions are derived from the LDA electronic structure. For instruction, 
Fig.~\ref{fig:fermi} shows the LDA Fermi surface of NdNiO$_2$ with its three distinct sheets,
namely the larger Ni-$d_{x^2-y^2}$ dominated sheet as well as the pocket sheets around 
$\Gamma$ and around A. Because of the strong downfolding to three orbitals, the disentangling 
procedure~\cite{sou01} is employed and the twofold dispersion closest to the Fermi surface 
is fixed in an inner energy window. The
result of this Wannier construction is displayed in Fig.~\ref{fig:wann}. The overall
Wannier dispersion agrees well with the LDA one, and the fatband analysis clearly 
identifies the Ni-$d_{z^2}$, Ni-$d_{x^2-y^2}$ and SD bands. Solely the 
dispersions/fatbands around the A point are hard to exactly align within the present 
orbital setting. This can be easily understood from the fact that the occupied part of 
this very region has sizable contribution from Ni-$t_{2g}$. However again, from 
our more general DFT+sicDMFT description, the very details around the A point are 
not crucial for the IL nickelate physics. It also becomes clear from Fig.~\ref{fig:wann},
that other DFT bands in the energy window $\sim[-2,-1]$\,eV are not included in the
minimal desription, and therefore it does not resemble a true ``monotonic'' low-energy 
Hamiltonian. However this is tailored to the present problem, and ``backed-up'' by the
full-fletch DFT+sicDMFT study: those left out DFT bands are of dominant Ni-$t_{2g}$
character and remain to a good approximation filled spectators throughout 
the phase diagram (see also comment in section~\ref{sumdis}).

But importantly note that the present SD orbital does not only describe the electron 
pockets at $\Gamma$ and A. It also has nonzero orbital weight on the Ni-$e_g$ dominated
dispersions, mostly on the Ni-$d_{z^2}$ band. This fact is generally very relevant to 
appreciate the role of Ni-$d_{z^2}$, and to understand the following results with
doping.

The real-space Wannier orbitals are shown in the bottom part of Fig.~\ref{fig:wann}.
For Ni-$e_g$, they resemble the expected appearance. 
Substantial leaking of Ni-$d_{x^2-y^2}$ onto inplane oxygen sites is a natural
outcome of the strong downfolding. Of course by construction, here the SD orbital 
cannot resemble an atomic-like orbitals Since not based on a highly-localized 
viewpoint, its spread is  comparatively large and the Wannier centre is inbetween Ni
and Nd. It as contributions from Nd, Ni and O and furthermore breaks the full-cubic 
symmetry. This orbital is important for its 'stand-alone' features to Ni-$e_g$.

Let us conclude this subsection by providing the Wannier values for the local 
single-particle terms $\varepsilon^{\rm loc}$ and for near hoppings $t$ in 
Tab.~\ref{tab:hopp}.  The nearest-neighbor hopping between Ni-$d_{x^2-y^2}$ and 
SD, here notably included in $\varepsilon^{\rm loc}$, is indeed zero, but there 
is a sizable one between Ni-$d_{z^2}$ and SD.

\subsection{RISB solution\label{sec:results}}
The paramagnetic RISB solution of the minimal Hamiltonian $H_{\rm min}$ in the 
stoichiometric half-filled case and with finite doping is discussed. Technically
speaking, the present multiorbital Hamiltonian is solved in a 'cluster-RISB' fashion,
since the onsite Ni-$e_g$ effects and the intersite effects between Ni-$e_g$ and the
SD orbital are treated on equal footing.

\subsubsection{Stoichiometry}
\begin{figure}[t]
\includegraphics*[width=8.5cm]{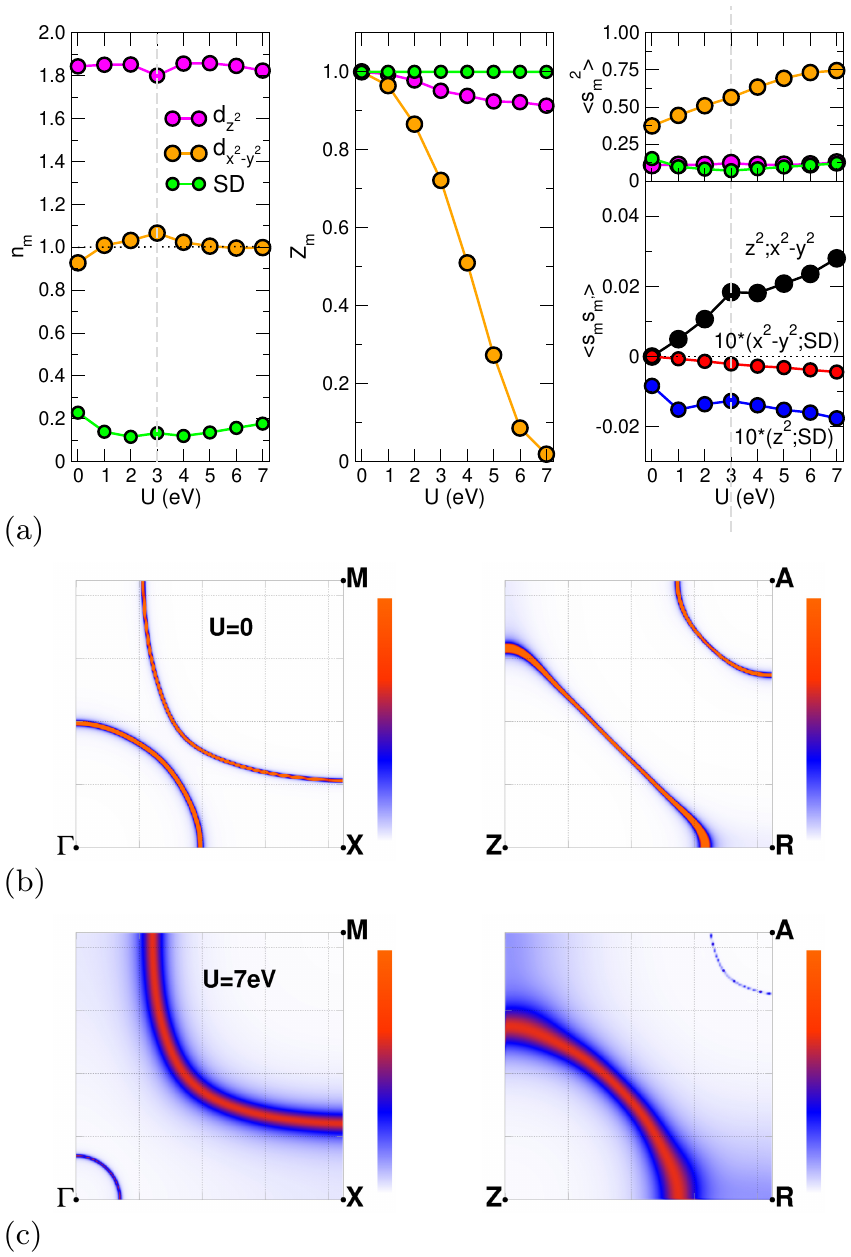}
\caption{(color online) Minimal-Hamiltonian properties at half filling for 
$\mu_{\rm SD}=U/2$.
(a) Selected quantities for increasing $U$. The Hund's
exchange is chosen $J_{\rm H}=U/3$ for $U<3$ and fixed to $J_{\rm H}=1$\,eV for 
$U\ge 3$\,eV (dashed gray line separates the regimes). 
From left to right: orbital occupation $n_m$, orbital-resolved QP weight
and local squared-spin moment (top) as well as local spin-spin correlation (bottom).
(b) Non-interacting Fermi surface for $k_z=0$ (left) and $k_z=1/2$ (right).
(c) Interacting Fermi surface for $U=7$\,eV.}
\label{fig:modstoich}
\end{figure}

Pristine NdNiO$_2$ resembles the half-filled $n=3$ case of the three-orbital description. 
At $U=0$, the orbital fillings read $\{n_{z^2},n_{x^2-y^2},n_{\rm SD}\}=\{1.84,0.93,0.23\}$,
i.e. the Ni-$d_{x^2-y^2}$ orbital lies somewhat below true half filling. 
Figure~\ref{fig:modstoich}a shows the evolution of key quantities of the low-energy 
electronic structure with increasing the Hubbard $U$. Be aware that due to the strong
downfolding, the lack of charge self-consistency as well as the RISB treatment, the
tailored $U$ value surely differs from $U=10$\,eV of the DFT+sicDMFT study. 

As $U$ grows from zero, charge transfers occur such to establish a truly half-filled 
Ni-$d_{x^2-y^2}$ orbital in an orbital-selective Mott transition scenario. For $U=7$\,eV, 
the fillings read $\{n_{z^2},n_{x^2-y^2},n_{\rm SD}\}=\{1.83, 1.0, 0.17\}$ and the 
quasiparticle (QP) weight $Z$ (i.e. the inverse effective mass) of the inplane orbital 
reads $Z_{x^2-y^2}=0.02$, i.e. is close to zero. In the following, we will understand the
electronic states at $U=7$\,eV as the low-energy equivalent to the more 
comprehensive DFT+sicDMFT picture at $U=10$\,eV.
The QP weight $Z_{z^2}$ remains close to unity due to the rather large filling. The 
localization of the Ni-$d_{x^2-y^2}$ electron is accompanied by the build-up of a
corresponding local squared spin moment $\langle s^2\rangle$ which saturates at the
limiting value $s(s+1/2)=3/4$. The Ni-$e_g$ spin-spin correlations are surely positive
because of Hund's first rule, and increase with the localization degree of 
Ni-$d_{x^2-y^2}$.
On the other hand, the intersite spin-spin correlations of $\{$Ni-$e_g$,SD$\}$ kind are 
negative and much smaller in magnitude. The $\{$Ni-$d_{z^2}$,SD$\}$ spin correlations 
turn out still larger than the $\{$Ni-$d_{x^2-y^2}$,SD$\}$ ones. This again proves that 
a {\sl direct} spin-spin coupling between the latter orbitals is weak.
\begin{figure}[t]
\includegraphics*[width=8.5cm]{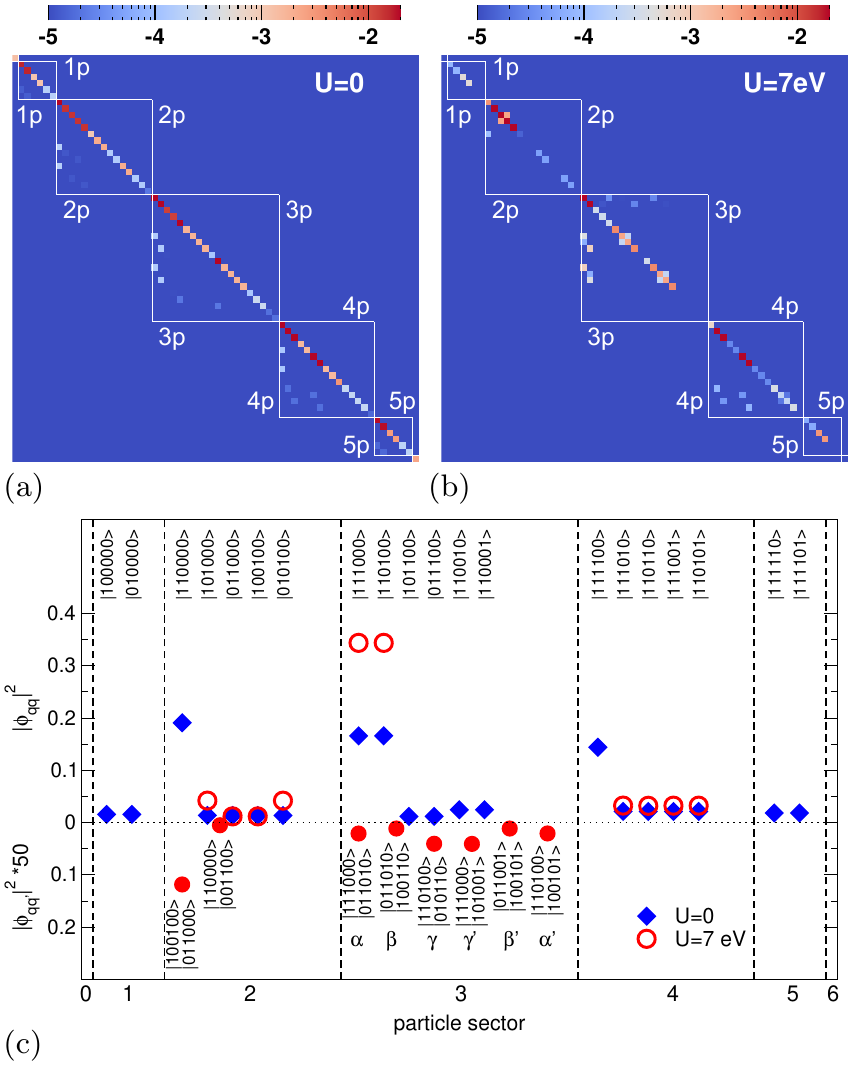}
\caption{(color online) Local-state behavior of the minimal Hamiltonian via the 
slave-boson weights $|\phi_{qq'}|^2$ connecting Fock states $q,q'$.
(a,b) Weight matrix in $q,q'$-space covering the $N=0,\ldots,6$ particle sectors $N$p
for weights $>10^{-5}$:
(a) $U=0$, and (b) $U=7$\,eV. (c) Dominant diagonal weights 
(along positive axis) for $U=0$ and for $U=7$\,eV, 
and dominant off-diagonal weights (along negative axis) within
each particle sector for $U=7$\,eV (filled circles).}
\label{fig:modsb}
\end{figure}

In view of the dichotomy between itinerant and localized behavior in strongly correlated
materials, let us further focus on Fermi surfaces and local states. The non-interacting 
Fermi surface in Fig.~\ref{fig:modstoich}b shows the electron pockets around $\Gamma$ and 
A, as well as the Ni-$d_{x^2-y^2}$-dominated sheet in the $k_z=0,1/2$ planes. Note the
closing of the latter sheet along Z$-$R as a difference to conventional
cuprate fermiology (see Ref.~\onlinecite{kar20} for a direct comparison).
In the strongly interacting case, the $\Gamma$ pocket is shrunk and since the QP at 
$U=7$\,eV is not yet exactly zero, the strongly-renormalized Ni-$d_{x^2-y^2}$ sheet still 
contributes to the Fermi surface (yet now slightly enlarged due to the exact half filling). 
But in the $k_z=1/2$ plane, the near-Mott-insulating sheet apparently now bends 
electron-like. The warping close to $k_z=1/2$ is therefore further strengthened with strong 
correlations. We will not further comment on the small-A-pocket features
because of possible Wannier-construction artifacts in that region for $U\ne 0$.

The slave-boson amplitudes $\phi_{qq'}$ connecting states $q,q'$ in the RISB formalism 
provide useful insight in the local-state behavior for different degrees of electronic 
correlation. Let us first remind about the corresponding Hilbert space. There are a 
total of $2^6=64$ (Fock) states available for our three-orbital system, with each 
particle sector $N=0,\ldots,6$ contributing $N_q=6!/(N!(6-N)!)$ states. Thus, if coupling
between {\sl all} states is allowed, the total number of $\phi_{qq'}$ amounts to
$64\times64=4096$ for a three-orbital problem.
However here, we only study normal-state properties and therefore exclude couplings
between different particle sectors. A RISB implementation for pairing problems permitting
also such couplings has been put forward by Isidori and Capone~\cite{isi09}.
Without pairing, one ends up with $\sum_N N_q^2=924$ slave-boson amplitudes (without using
symmetries) in the RISB calculation. The squared amplitudes of the relevant part of those 
are depicted in Fig.~\ref{fig:modsb} for the non-interacting case and for $U=7$\,eV. 
The values $|\phi_{qq'}|^2$ may be interpreted as the weight for finding the system in a 
quantum state characterized by $q,q'$ to occur, since $\sum_{qq'}|\phi_{qq'}|^2=1$ holds. 
There is of course the option to transfer from a Fock basis to a multiplet 
basis~\cite{pie18}, but due to the specific itinerant inter-site structure of our 
Hamiltonian we remain in the Fock basis for a straightforward analysis.

Figures~\ref{fig:modsb}a,b first show that for sizable magnitudes the weight matrix in 
$q,q'$-space is mostly diagonal. Especially without interaction, quantum entanglement 
between unlike states is rather implausible. But also along the diagonal, 
the weights are much more distributed for $U=0$, just because there is no 
interaction-driven state selection but only statistics of a Fermi gas. 
Since describing half filling, for $U=7$\,eV the fluctuations to the 0,1 and 5,6 
particle sectors is significantly suppressed. The truly relevant diagonal Fock 
states are easily selected (see Fig.~\ref{fig:modsb}c), using the notation 
\begin{equation}
|q\rangle=|\uparrow_{z^2}\downarrow_{z^2}\;
\uparrow_{x^2-y^2}\downarrow_{x^2-y^2}\;\uparrow_{\rm SD}\downarrow_{\rm SD}\rangle\;.
\end{equation}
Nonsurprisingly the largest weight stems from the $|111000\rangle$ and $|110100\rangle$
states in the 3-particle sector, i.e. two electrons in Ni-$d_{z^2}$ and one in
Ni-$d_{x^2-y^2}$. Of course, doubly-occupied (sub-)states in Ni-$d_{x^2-y^2}$, though
appreciated without interactions, are highly suppressed at large $U$. There are still
four relevant diagonal states for $U=7$\,eV in the 2- and 4-particle sector, respectively.
The ones in the 2-particle sector display Hund's first rule within Ni-$e_g$. The ones
in the 4-particle sector exhibit fully occupied Ni-$d_{z^2}$ as well as singly-occupied 
Ni-$d_{x^2-y^2}$ and SD, but importantly, without spin-alignment 
differentiation for the $|\phi_{qq'}|^2$ magnitude. In other words, without charge
fluctuation in Ni-$d_{z^2}$ there is no exchange-favored discrimination of 
$\{$Ni-$d_{x^2-y^2}$,SD$\}$ spin states. 
\begin{figure}[t]
\includegraphics*[width=8.5cm]{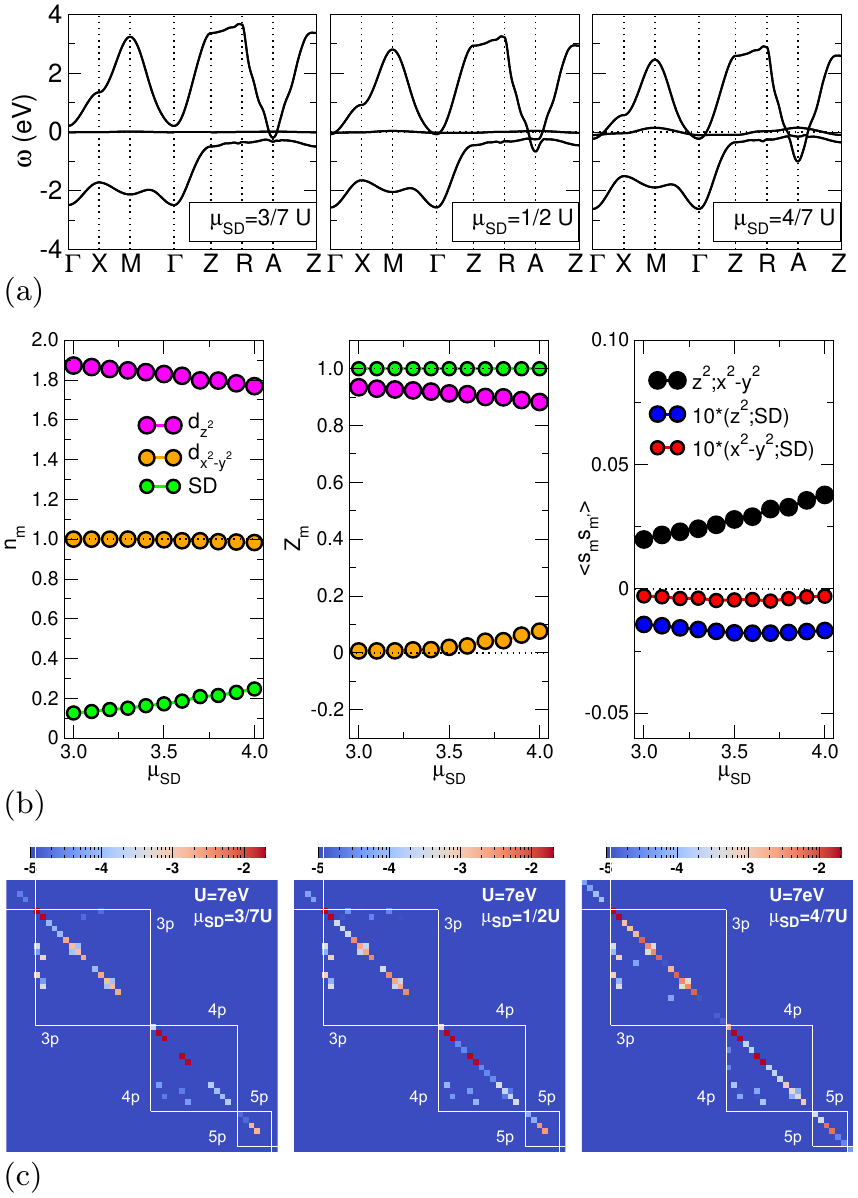}
\caption{(color online) Effect of $\mu_{\rm SD}$ on the electronic states
for $U=7$\,eV.
(a) QP dispersion. (b) Orbital-resolved occupations, QP weights and local
spin-spin correlations. (c) Slave-boson weight matrix $|\phi_{qq'}|^2$
in higher particle-sectors.}
\label{fig:sd}
\end{figure}

Let us finally turn to the off-diagonal slave-boson weights, of which there are a few
in the strongly interacting regime with sizable magnitude (cf. Figs.~\ref{fig:modsb}b,c), 
but still much smaller than the diagonal ones. There is a prominent $\phi_{qq'}$ in the 
2-particle sector, connecting the two Ni-$e_g$ states $|100100\rangle$ and
$|011000\rangle$. This clearly marks the relevant singlet $S=0, S_z=0$ and 
triplet $S=1, S_z=0$ entanglement in the $e_g$ manifold. The respective twofold 
off-diagonal weights $\alpha^{(')}$, $\beta^{(')}$ and $\gamma^{(')}$ in the 
3-particle sector are more complicated, but most interesting. For clarity, let us
enumerate them in words:
\begin{itemize}
\item[$\alpha^{(')}$:] scattering between doubly Ni-$d_{z^2}$, spin-up(down)
Ni-$d_{x^2-y^2}$ and spin-up(down) Ni-$d_{z^2}$, spin-down(up) Ni-$d_{x^2-y^2}$,
spin-down(up) SD.
\item[$\beta^{(')}$:] scattering between singlet/triplet Ni-$d_{z^2}$, Ni-$d_{x^2-y^2}$
and spin-up/down SD.
\item[$\gamma^{(')}$:] scattering between doubly Ni-$d_{z^2}$, spin-up(down)
Ni-$d_{x^2-y^2}$ and spin-up(down) Ni-$d_{z^2}$, spin-up(down) Ni-$d_{x^2-y^2}$,
spin-down(up) SD.
\end{itemize}
Thus these weights provide the desired spin differentiation in the scattering processes
between the three orbitals, relevant for paving the way toward establishing a proper
Kondo(-lattice) picture. The weights accordingly scale as 
$\gamma^{(')}>\alpha^{(')}>\beta^{(')}$, which importantly provides further quantitative
evidence for a Hund-assisted Kondo scenario via Ni-$d_{z^2}$, leading to AFM exchange
between localized Ni-$d_{x^2-y^2}$ and the SD band.

To conclude this discussion at stoichiometry, we want to comment on the role of the 
potential shift $\mu_{\rm SD}$, which was set so far to $\mu_{\rm SD}=U/2$. 
Figure~\ref{fig:sd} depicts the modifications in the electronic states 
with varying $\mu_{\rm SD}$ to smaller and larger values. Generally, the effects
from changing $\mu_{\rm SD}$ by reasonable amounts are comparatively weak.
The given potential shift defines the energy location of the SD band, thus a 
smaller(larger) value shifts it upwards(downards) with respect to $\mu_{\rm SD}=U/2$ 
(see Fig.~\ref{fig:sd}a). 
Understandably, a larger $\mu_{\rm SD}$ leads a stronger filling of the SD band and
a decrease of correlation strength in Ni-$d_{x^2-y^2}$ (see Fig.~\ref{fig:sd}b).
The growth of the $\{$Ni-$d_{z^2}$, Ni-$d_{x^2-y^2}$$\}$ spin-spin correlation with 
increasing $\mu_{\rm SD}$ in Fig.~\ref{fig:sd}c may be attributed to the less-filled 
Ni-$d_{z^2}$ in this situation, strengthening the Ni-$d_{z^2}$ moment. The $\mu_{\rm SD}$
variation renders furthermore the coupling between itinerant and local degrees
of freedom obvious. Figure~\ref{fig:sd}c shows for instance the growth of an off-diagonal
$|\phi_{qq'}|^2$ near the diagonal at the upper end of the 4-particle sector. This
amounts to the coupling of states $\{|011011\rangle$, $|100111\rangle\}$, 
describing scattering between singlet/triplet Ni-$e_g$ and doubly-occupied SD. 
Thus, the growth of SD occupation from a corresponding band that sinks below the Fermi 
level also leads to an increase of $\phi_{qq'}$ amplitudes associated with a completely 
filled SD state.

The observed $\mu_{\rm SD}$ variations provide confidence that at stoichiometry, the 
setting $\mu_{\rm SD}=U/2$ indeed proves reasonable. For instance, the existence of
the $\Gamma$ electron-pocket Fermi surface is a robust feature of the DFT+sicDMFT
study (and also reported from many other theoretical works).

\subsubsection{Hole doping}
The experimental doping with Sr and accordingly the replacment of Nd$^{3+}$ by
Sr$^{2+}$ leads to the doping of holes into the nickelate compound. In 
Ref.~\onlinecite{lec20} we studied this doping process by a DFT+sicDMFT 
treatment of realistic supercells for Nd$_{1-x}$Sr$_x$NiO$_2$ at $T=193$\,K. 
It was found that for $x=0.125$, while both Ni-$e_g$ orbital gain holes (though
Ni-$d_{z^2}$ slightly more), the lowest-energy response comes mainly 
from Ni-$d_{x^2-y^2}$. Whereas for $x=0.25$, the number of holes in Ni-$d_{z^2}$ is 
quite large and the lowest-energy response is twofold but dominated by Ni-$d_{z^2}$.
This is apparently not in line with a cuprate-like, single-orbital hole doping of
Ni-$d_{x^2-y^2}$.

In the following, hole doping is investigated in more detail from the low-energy 
perspective within the minimal-Hamiltonian approach. It amounts to change the total
filling to $n=3-\delta$ with $\delta>0$. This approach surely misses some effects of 
the realistic Nd substitution by Sr. But it should be geared to unveil the key 
qualitative features with doping. For instance, one (also electrostatic-supported) 
effect of true Sr doping is the shifting of the $\Gamma$ electron pockets into the 
unoccupied region~\cite{zha20,bot20,lec20,leo20}. In the minimal-Hamiltonian scheme, 
part of the corresponding underlying mechanism surely goes into the value of 
$\mu_{\rm SD}$. We hence used the two values $\mu_{\rm SD}=3/7U,U/2$ for comparison 
and to truly include that upward shifting qualitatively correct.
\begin{figure}[t]
\includegraphics*[width=8.5cm]{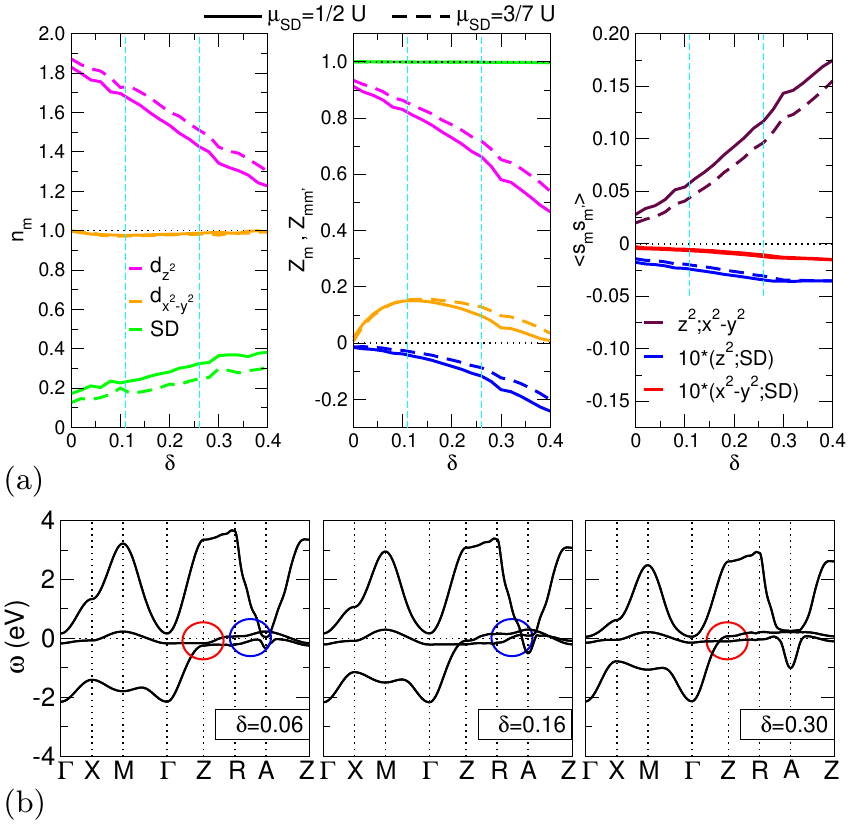}
\caption{(color online) Evolution of NdNiO$_2$ many-body electronic structure with
hole doping $\delta$, approached by the three-orbital Hamiltonian at $U=7\,$eV.
(a) From left to right: orbital-resolved occupations, QP weights and local
spin-spin correlations for $\mu_{\rm SD}=3/7U$ (dashed lines) and
$\mu_{\rm SD}=U/2$ (solid lines). Cyan dashed lines mark the occurrence of
Fermi-surface reconstructions (see text).
(b) QP dispersions at selected $\delta$ for $\mu_{\rm SD}=3/7U$.
Places for the Fermi-surface reconstructions are highlighted by
blue ellipse (for occurring at $\delta=0.11$) and by red ellipse
(for occurring at $\delta=0.26$).}
\label{fig:hdop}
\end{figure}

Figure~\ref{fig:hdop} summarizes the key changes with hole doping from the 
minimal-Hamiltonian perspective for $U=7$\,eV. Notably, the differences between 
$\mu_{\rm SD}=3/7U,U/2$ are small and mostly quantitative. The first surprising 
(though already DFT+sicDMFT suggested) observation is the very weak change of the 
Ni-$d_{x^2-y^2}$ orbital filling with increasing $\delta$ (cf. Fig~\ref{fig:hdop}a).
It remains very near half filling up to the investigated limiting value $\delta=0.4$.
In fact starting from stoichiometry, $n_{x^2-y^2}$ first decreases slightly until
$\delta\sim 0.1$, and increases afterwards progressively toward half filling. 
On the other hand, $n_{z^2}$ decreases steadily with $\delta$. The second surprise 
is the {\sl increase} of the SD filling with hole doping. As one associates with SD 
the electron pockets, one naively expects a $\delta$-induced depopulation of that 
orbital. However, one has to recall the important fact that the SD state hybridizes 
substantially with Ni-$d_{z^2}$ over a large energy region (see right 
bottom of Fig.~\ref{fig:wann}). With large $U$ on Ni, the $e_g$ electrons try hard to 
escape to the SD state, if they cannot achieve complete localization. Hence there is an
correlation-induced change of the hybridization, here apparently activated for 
hole doping, such that the SD state increases its weight on the deep-lying 
Ni-$d_{z^2}$-dominated band and more electrons can benefit from the favorable 
SD-orbital location. 
\begin{figure}[t]
\includegraphics*[width=8.5cm]{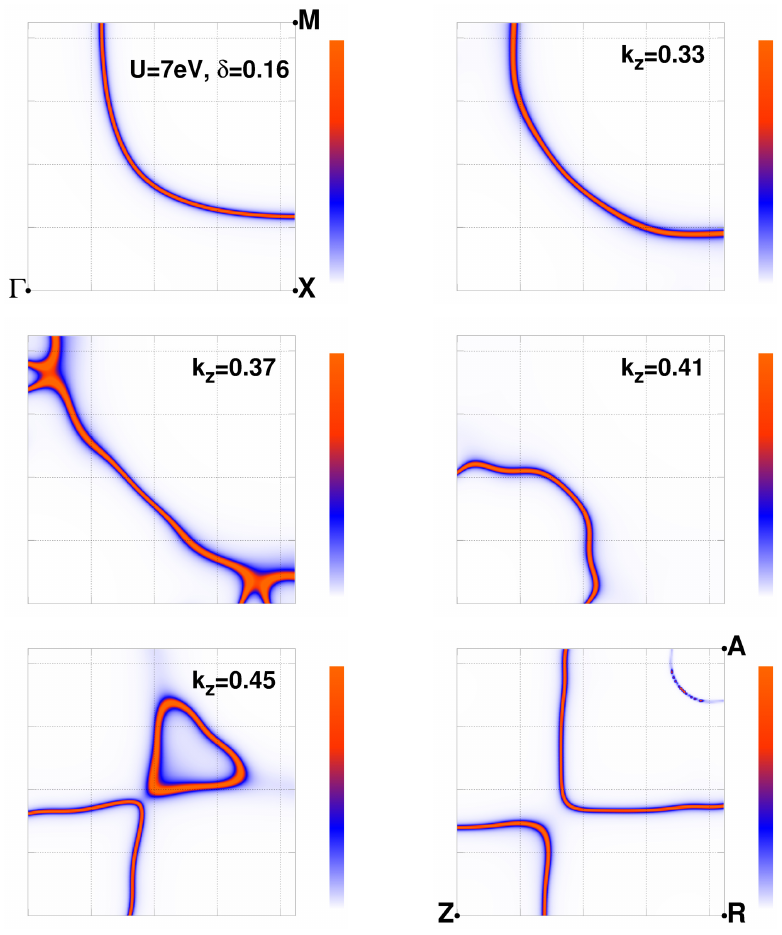}
\caption{(color online) The interacting Fermi surface along $k_z$ at $\delta=0.16$, 
for $U=7\,$eV and $\mu_{\rm SD}=3/7U$.}
\label{fig:ifs}
\end{figure}
\begin{table}[t]
\begin{tabular}{C{1.6cm}|C{1.9cm}|C{1.45cm} C{1.45cm} C{1.45cm} }
  orbital  & Fermi sheets & $\delta=0.06$ & $\delta=0.16$ & $\delta=0.30$\\ \hline
                      & around Z   &    moderate   &   strong    &  $-$     \\
\bw{Ni-$d_{z^2}$}     & around A   &     $-$       &   moderate  &  weak    \\ \hline
                      & around Z   &    strong     &   moderate  &  $-$     \\
\bw{Ni-$d_{x^2-y^2}$} & around A   &     $-$       &   moderate  &  strong  \\
\end{tabular}
\caption{Qualitative orbital-character strength on the interacting Fermi-surface sheets
in the $k_z$=1/2 plane for selected hole dopings. 
See bottom right of Fig.~\ref{fig:ifs} for a visualization of both Fermi sheets.}
\label{tab:mix}
\end{table}
This mechanism is proven by $Z_{z^2,\rm SD}$ being the only sizable off-diagonal 
QP weight, and its absolute value steadily increasing with $\delta$ 
(see bottom of Fig~\ref{fig:hdop}b). This off-diagonal $Z$ monitors the outlined 
inter-site hybridization modification between Ni-$d_{z^2}$ and SD state.
Thus, the strong doping of Ni-$d_{z^2}$ is triggered by the direct hole creation
from $\delta$, as well as additional charge transfer to the SD state. All this
apparently being energetically more favorable than putting more holes into 
Ni-$d_{x^2-y^2}$.

The diagonal QP weights $Z_m$ behave then as expected with $\delta$, i.e. 
$Z_{z^2}$ decreasing and $Z_{x^2-y^2}$ exhibiting a non-monotonic behavior 
with a maximum around $\delta\sim 0.1$. Note again that the Ni-$d_{x^2-y^2}$
correlation strength at the large doping of $\delta\sim 0.4$ becomes comparable 
to the one at half filling. Expectedly, the spin-spin correlation between the
Ni-$e_g$ orbitals increases with $\delta$ due to the enhancement of the 
Ni-$d_{z^2}$ localization.

The described evolution in fillings and many-body observables are accompanied by
obvious changes of the interacting Fermi surface, shown in Fig.~\ref{fig:hdop}b.
For $\delta<0.11$, the fermiology in the $k_z=1/2$ plane is basically equivalent
to the stoichiometric scenario, i.e. showing an electron-like pocket around $Z$ 
(see right part of Fig.~\ref{fig:modstoich}c). Yet for $\delta>0.11$, there is
a Lifshitz transition with an emerging second pocket around A. This pocket should
not be confused with the original small SD-based electron pocket. The Lifshitz 
transition is realized by a doping- and correlation-induced shift of the rather 
flat Ni-$d_{z^2}$-dominant dispersion in the $k_z=1/2$ plane onto the Fermi level. 
Note that already at stoichiometry, the enhanced proximity of the latter dispersion 
to $\varepsilon_{\rm F}$ compared to IL cuprates has been documented within 
LDA~\cite{lec20}, and was also revealed within DFT+U~\cite{cho20}.

Hence the Fermi surface reconstructs at $\delta\sim 0.11$, giving rise to the 
topology shown in Fig.~\ref{fig:ifs} for $\delta=0.16$. Going along $k_z$, the 
original Ni-$d_{x^2-y^2}$ hole pocket bends over toward an electron pocket,
shown for $k_z=0.41$. At $k_z=0.45$ a new heart-shaped hole pocket starts out
along the nodal line. It opens up to a large hole sheet around A at $k_z=1/2$.
The nature of the novel larger pocket around A is of mixed Ni-$e_g$ kind. There 
is still another Lifshitz transition at larger $\delta$. For $\delta<0.26$ the 
van-Hove singularity at $Z$ lies below the Fermi level, and above for $\delta>0.26$. 
Therefore, the $k_z=1/2$-pocket around $Z$ vanishes for hole doping beyond 
$\delta=0.26$. It is surely tempting to relate these reconstructions with the 
experimental phase boundaries of superconductivity~\cite{li20,zen20}. 

\begin{figure}[b]
\includegraphics*[width=8.5cm]{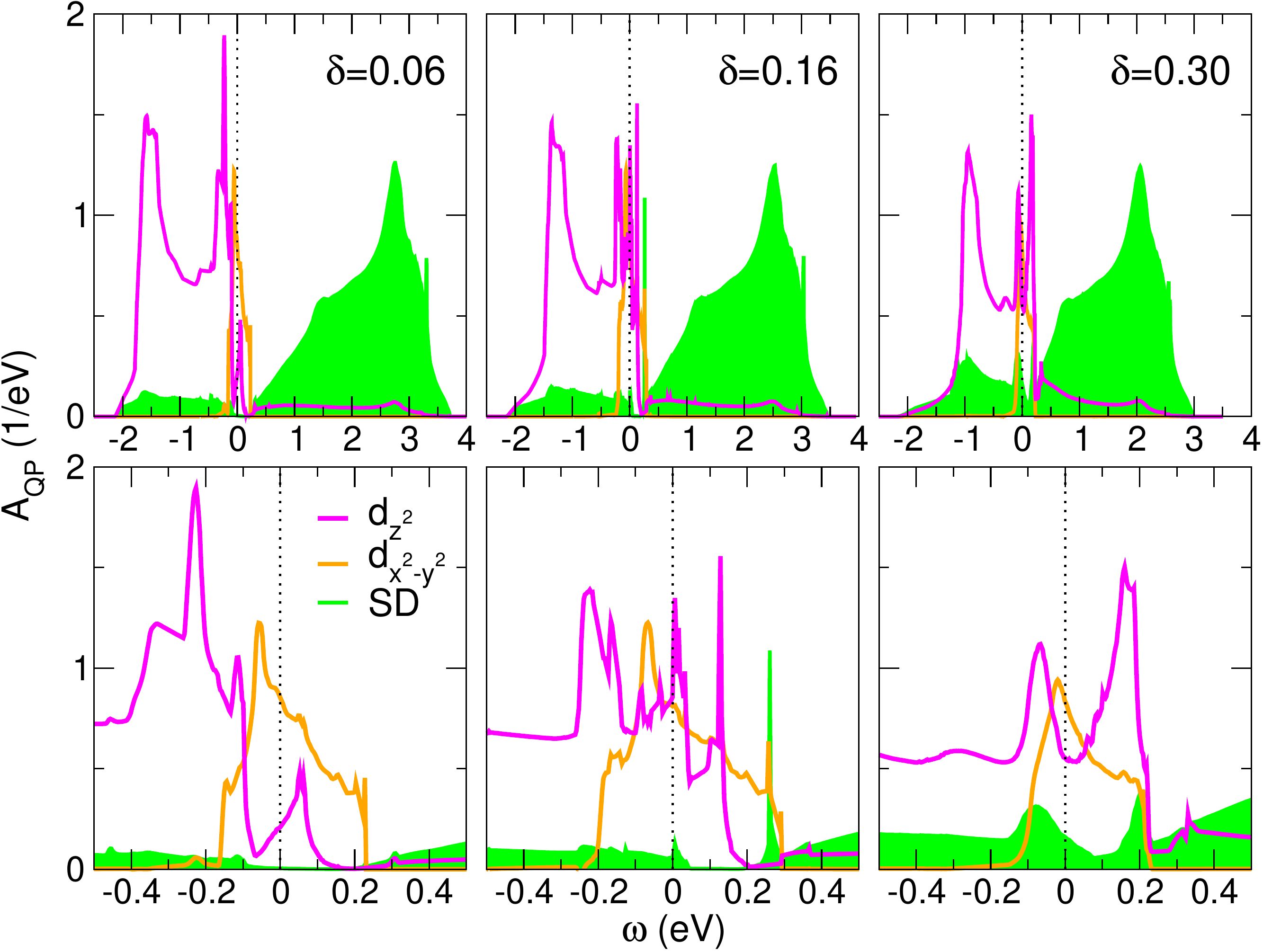}
\caption{(color online)
Orbital-resolved {\bf k}-integrated QP dispersion for selected hole dopings $\delta$ 
based on the minimal-Hamiltonian approach. Top (Bottom): wide (narrow) energy window.
Since there are no Hubbard bands in mean-field RISB, note that each orbital spectrum 
is multiplied by the respective $Z_m$ to account for the proper relative QP weight.}
\label{fig:dopdos}
\end{figure}

The {\bf k}-integrated orbital-resolved QP spectral functions shown in 
Fig.~\ref{fig:dopdos} for selected dopings, underline the scenario we just discussed:
For $\delta=0.06$, the Ni-$d_{z^2}$ weight at the Fermi level is still small,
whereas for $\delta=0.16$ it is larger than Ni-$d_{x^2-y^2}$ and peaks at 
$\varepsilon_{\rm F}$. The latter intermediate region therefore displays the strongest 
susceptibility for Ni-$d_{z^2}$-driven instabilities in a highly-localized 
Ni-$d_{x^2-y^2}$ background. For $\delta=0.30$, the Ni-$d_{z^2}$ weight close
to $\varepsilon_{\rm F}$ is again smaller, since the Fermi level is located in a
``pseudogap'' of Ni-$d_{z^2}$ weight. This structure is well understood from 
Ni-$d_{z^2}$ occupation of states with small $k_z$ and and unoccupied states with 
$k_z$ closer to 1/2, due to the doping-dependent QP band shifts.

Yet a further complexity issue has to be discussed. Upon increasing $\delta$ and running
through the described Fermi surface reconstructions, the character partition between
Ni-$d_{z^2}$ and Ni-$d_{x^2-y^2}$ on the (near-)$k_z=1/2$ Fermi sheets also varies. Reason
is that the original bands of both of these orbital characters sweep through each other in 
that region of the Brillouin zone with doping (see Fig.~\ref{fig:modstoich}b). 
Hence a nontrivial mixing of characters due to correlation-modified hybridizations 
takes place. Table~\ref{tab:mix} summarizes the basic information in simple terms for 
selected dopings in the three distinct regions.
It is seen that the intermediate doping region $0.11<\delta<0.26$ displays the most 
intriguing case with the strongest multiorbital character of the Fermi surface. Surely note
that the issue of mixed Fermi-surface character does not arise in the $k_z$=0 plane:
The $\Gamma-$M pocket is of perfect Ni-$d_{x^2-y^2}$ kind throughout the studied 
hole-doping region.

In fact, recent experimental Hall data in the normal state is interpreted as going
from an electron- to a hole-like transport scenario with doping~\cite{li20,zen20}.
This is in line with our findings, since for small $\delta$, transport should mainly
arise from the vanishing SD electron pockets and the electron-like bended region around 
$Z$ which has mixed Ni-$e_g$ character (cf. Tab.~\ref{tab:mix}). But for $\delta>0.11$, 
the hole pocket around $A$ appears, showing strong Ni-$d_{z^2}$ weight for intermediate 
doping. 
\begin{figure}[t]
\includegraphics*[width=8cm]{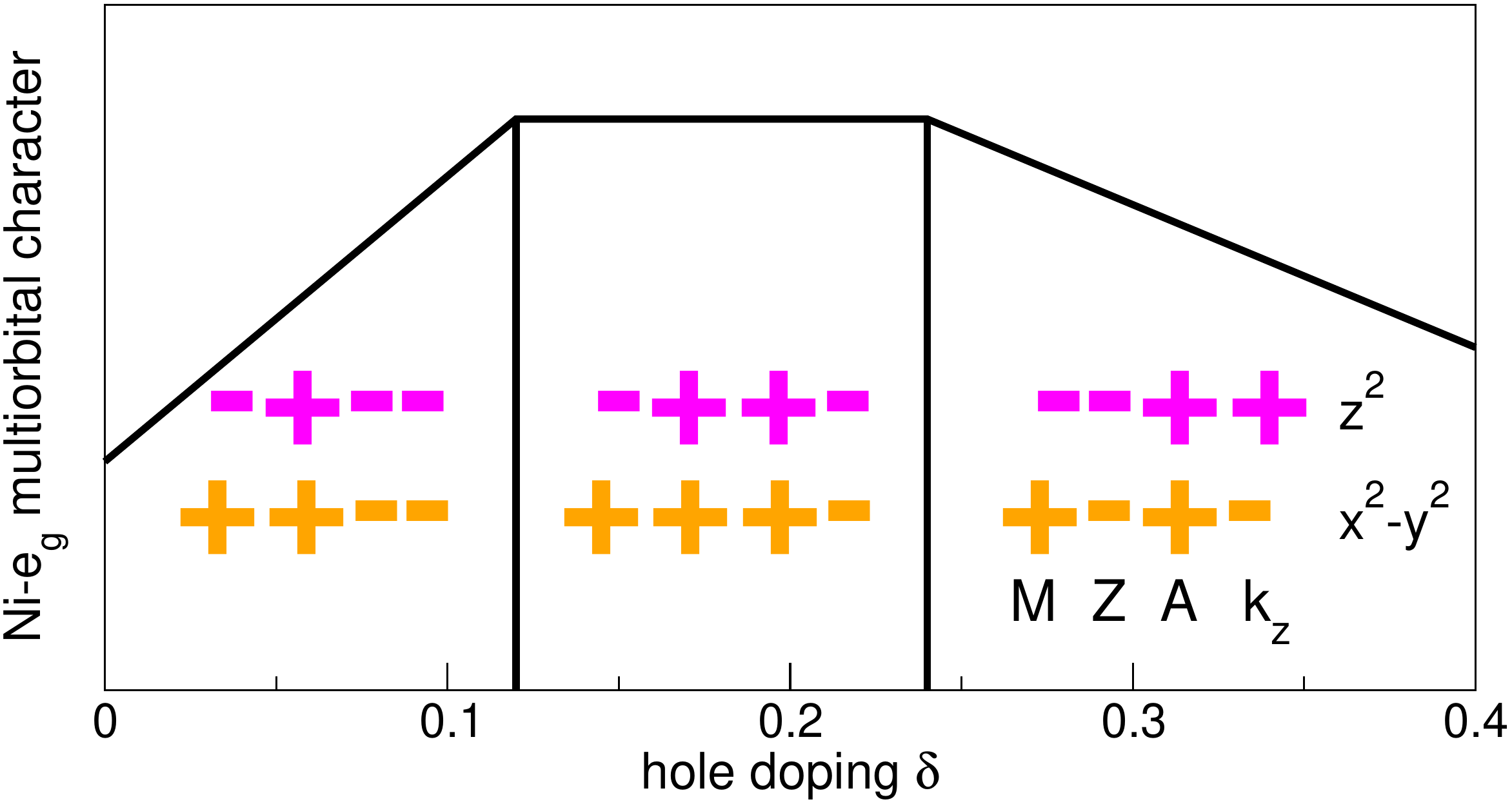}
\caption{(color online) Schematic sketch of the Ni-$e_g$ multiorbital character with 
$\delta$. The '$+$' symbols denote a Fermi-sheet participation of $x^2-y^2$ or $z^2$,
respectively, around M, Z, A or along $k_z$ in the Brillouin zone.}
\label{fig:dia}
\end{figure}

Last but not least, the present results allow us moreover to also provide an explanation 
for the puzzling weakly-insulating behavior for $x>0.25$ in experiment~\cite{li20,zen20}. 
First, the SD electron pockets are above the Fermi level at larger hole doping. 
Second, Ni-$d_{x^2-y^2}$ remains essentially half-filled and localized for large $\delta$, 
the $k_z=0$ Fermi sheet is Ni-$d_{x^2-y^2}$ dominated and so is the only remaining 
in the $k_z=1/2$ plane (see Tab.~\ref{tab:mix} for $\delta=0.30$). Thus transport from 
that orbital sector is by any means very weak at larger $\delta$.  
And third, the only remaining Fermi sheet with strong Ni-$d_{z^2}$ character for 
large $\delta$ lies {\sl along} $k_z$, i.e. between $\Gamma-$Z 
(see Fig.~\ref{fig:hdop}c for $\delta=0.30$). But that sheet will mainly account for 
transport in $c$-axis direction of a Nd$_{1-x}$Sr$_x$NiO$_2$ crystal. But transport 
measurements have so far been performed on thin films.
Putting all this together, inplane conductivity at hole dopings beyond the upper 
superconducting phase boundary is indeed expected very weak from our theoretical study.

A summary of the $\delta$-variation of the Ni-$e_g$ multiorbital character is 
schematically sketched in Fig.~\ref{fig:dia}, where the doping-dependent Fermi-surface 
sheets are counted with respect to their participation of either of the Ni-$e_g$ flavor. 
It is seen that the region of intermediate region of hole doping, interestingly rather 
well agreeing with the experimental region for superconductivity, is designated with 
the most pronounced multiorbital character.

\subsubsection{Comparison between electron and hole doping}
The dichotomy between electron and hole doping is an essential part of the high-$T_c$
cuprates physics~\cite{arm10}.
Therefore in the final result section, the opportunity is used to compare in basic
terms the electron-doped region within the minimal-Hamiltonian approach with the hole-doped
one. Experimentally, electron doping of NdNiO$_2$ has not been achieved yet. In analogy
to cuprates, it could formally being realized by e.g. replacing Nd$^{3+}$ with Ce$^{4+}$.
The theoretical calculations are straightforwardly extended to electron doping by fixing
the total particle number at $n=3-\delta$ with $\delta<0$. 

\begin{figure}[t]
\includegraphics*[width=8.5cm]{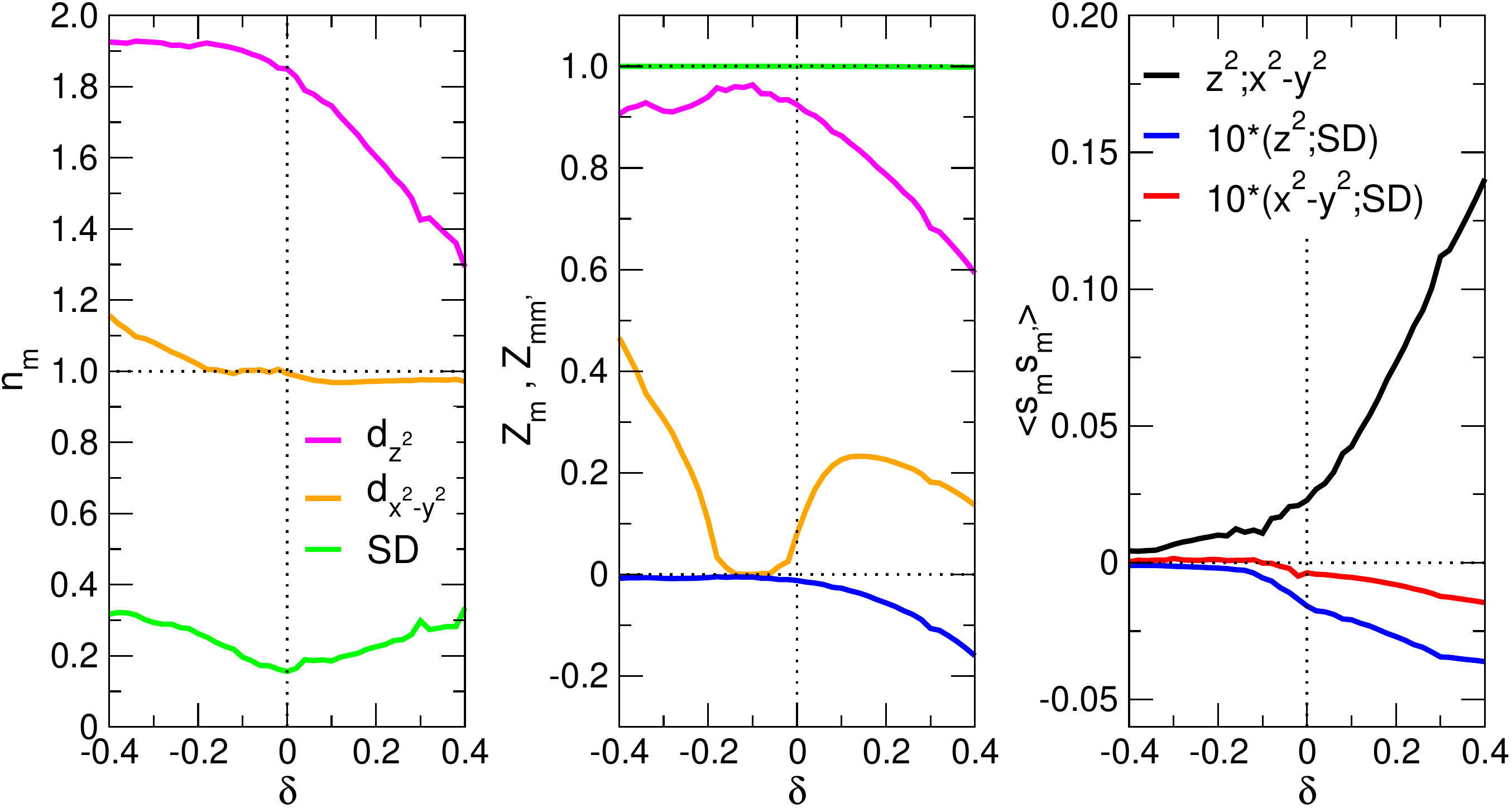}
\caption{(color online) Comparison of electron ($\delta<0$) vs. hole ($\delta>0$)
doping within the minimal-Hamiltonian picture for $U=6$\,eV. From left to right: 
orbital-resolved occupations, QP weights and local spin-spin correlations. 
The SD potential shift is set to $\mu_{\rm SD}=U/2$.}
\label{fig:ehdop}
\end{figure}

Figure~\ref{fig:ehdop} displays the comparison of both doping regimes. We lowered the 
Hubbard interaction to $U=6$\,eV, since the RISB numerics of the electron-doped region
turns out more challenging. Nonetheless, the hole-doped side behaves qualitatively identical
to the results for $U=7$\,eV.

Let us thus focus on the electron doping, which shows singular differences to hole 
doping. Starting with the QP weights, it is seen that for small $\delta$ the correlation
strenght in Ni-$d_{x^2-y^2}$ apparently {\sl increases} compared to half filling. 
After a minimum at $\delta\sim -0.1$, the QP weight $Z_{x^2-y^2}$ then strongly recovers
and grows up to the studied limiting value $\delta=-0.4$. The doping $\delta=-0.1$ apparently
marks the true orbital-selective Mott-transition point in the IL nickelate. The Ni-$d_{x^2-y^2}$
filling remains fixed at half filling for small $\delta$, but then also grows beyond the
latter transition point. Expectedly, the Ni-$d_{z^2}$ orbital has to accomodate even
more electrons with electron doping. Yet because of the large associated Coulomb-repulsion 
cost, it succeeds in not becoming completely filled, but can delegate some charge to
the remaining orbitals. Also without surprise, the SD orbital becomes further filled, too.
With electron doping, no sophisticated correlation-induced change of hybridization
has to be invoked in order to place additional electrons in the SD state. Correspondingly, 
the off-diagonal QP weight $Z_{z^2,\rm SD}$ is small for electron doping. Thus, albeit
original electron pockets of the SD band shift deeper into the occupied region, the
coupling at least to Ni-$d_{z^2}$ appears weaker. The Ni-$e_g$ spin-spin correlations 
are much smaller for electron doping, because of the even stronger Ni-$d_{z^2}$ filling.
Interestingly when adding electrons, the spin-spin correlations between Ni-$d_{x^2-y^2}$ 
and SD change sign from negative to positive at the critical point $\delta=-0.1$.

The electron-doped region might be interesting because of obviously stronger 
Ni-$d_{x^2-y^2}$ correlation at low doping and possibly intriguing competition between 
Kondo correlations and magnetic order. Superconducting tendencies of similar kind as 
on the hole-doped side are not expected, since electron-doping should not lead to comparable 
intricate Fermi-surface reconstructions as Ni-$d_{z^2}$ becomes too strongly filled.
However apparently, the electron-doped side supports a stronger Ni single-orbital picture
of Ni-$d_{x^2-y^2}$ kind. Thus despite the coexistence with the remaining SD band, similarities
to cuprates might be more pronounced with electron doping.

\section{Summary and Discussion\label{sumdis}}
Using comprehensive DFT+sicDMFT as well as an aligned minimal-Hamiltonian representation
solved within RISB, we were able obtain important insight into the very rich physics of 
NdNiO$_2$ in pristine condition and with finite doping.

Relevant features of a stoichiometric Kondo(-lattice) scenario at lower temperature are 
revealed, showing that a calculated sizable AFM Kondo coupling $J_{\rm K}\sim$ 120\,meV 
builds up onto the cooperation of Ni-$d_{x^2-y^2}$ , Ni-$d_{z^2}$ and the self-doping band. 
The Ni-$d_{z^2}$ and the SD band both take part in the screening of the Ni-$d_{x^2-y^2}$ 
spin, whereby the former orbital has a mediating role through a Hund-assisting mechanism.
The onset of significant Kondo correlations around $T\sim 60$\,K matches well with
the experimental $T=70$\,K for the beginning resistivity upturn~\cite{li19}. An 
interesting {\bf k}-selective Ni-$e_g$ hybridization via oxygen around $\Gamma$ deserves
further investigation.

A realistic minimal three-orbital Hamiltonian is advocated, that describes the low-energy
competition of both Ni-$e_g$ orbitals linked to a SD state that carries the effect of
remaining Nd$(5d)$, O($2p$), Ni-$t_{2g}$ and Ni$(4s)$. This Hamiltonian serves the goal of
providing a faithful representation of the key degrees of freedom of IL nickelate. Its 
canonical structure, i.e. two strongly interacting orbitals coupled to a 'stand-alone bath'
state, may prove useful for other nickelates or related problems. 

The system with hole doping $\delta$ is remarkably different to hole-doped cuprates. The 
Ni-$d_{x^2-y^2}$ occupation is only very weakly departing from half filling and the orbital
even regains correlation strength at larger $\delta$. On the other hand, the Ni-$d_{z^2}$ 
orbital eagerly collects holes. Also on the expense of the SD orbital, which 
counterintuitively gains electrons by a correlation-induced change of hybridization 
to Ni-$d_{z^2}$. While the QP weight for the latter orbital decreases with $\delta$, 
it remains in a weak-to-moderate correlation regime of $Z_{z^2}\sim 0.7$ at intermediate 
doping. Further key aspect at hole doping are two apparent Fermi-surface constructions that 
designate $0.11<\delta<0.26$ as the region with strongest multiorbital nature and 
entanglement. It is an orbital-selective(-kind) situation with highly-correlated, 
hardly-doped Ni-$d_{x^2-y^2}$ and still substantially-filled Ni-$d_{z^2}$ sharing 
two Fermi sheets in the $k_z=1/2$ plane of the Brillouin zone. This specific doping 
region agrees well with the two experimentally-determined ranges, i.e. 
$0.125<x<0.25$ by Li {\sl et al.}~\cite{li20} 
and $0.12<x<0.235$ by Zeng {\sl et al.}~\cite{zen20}, for the appearance of 
superconductivity in Nd$_{1-x}$Sr$_x$NiO$_2$.

Moreover, our realistic description may not only explain the experimentally observed
change from electron-like to hole-like transport in Hall measurements, but may
also provide reason for the weakly-insulating behavior found on {\sl both} sides of 
the supercondcuting region. In essence, Ni-$d_{x^2-y^2}$ remains
nearly localized for any reasonable hole doping, and the itinerant contribution of 
Ni-$d_{z^2}$ remains generally small for $\delta<0.11$ while contributing mostly to
$c$-axis transport for $\delta>0.26$. The latter should be masked in the so far
available thin-film studies of Nd$_{1-x}$Sr$_x$NiO$_2$.

Finally, theoretical electron doping leads to quite different characteristics than
hole doping. Most notably, even further correlation-strength enhancement within 
Ni-$d_{x^2-y^2}$ is found for small negative $\delta$, and the spin exchange between 
the latter and the SD orbital switches to ferro-like behavior. The 
predominant Ni-$d_{x^2-y^2}$ characteristics on the electron-doped side deserve 
future experimental investigations.

The present theoretical results raise several further questions. For instance, one 
wonders about the role of Ni-$t_{2g}$ states with hole doping, since those are also 
located in close distance to the Fermi level. However from our defect-supercell 
DFT+sicDMFT study~\cite{lec20}, which included all Ni$(3d)$ on equal footing, 
they become only weakly doped and their contribution to relevant low-energy 
physics appears minor up to described $x\sim 0.25$. 
An intuitive reason may be given by the fact that
already on the LDA level, Ni-$e_g$ is stronger dispersing than Ni-$t_{2g}$. At
strong coupling, with Ni-$d_{x^2-y^2}$ becoming half filled, the only appreciable 
itinerant Ni$(3d)$ degree of freedom to balance the electronic energy cost is thus
provided by Ni-$d_{z^2}$.
Second, if the IL nickelate superconductivity is based on multi-orbital Ni-$e_g$ 
processes and not on the so far singular 'cuprate mechanism', the question arises 
why stable pairing is not much more common in nickel-based transition-metal oxides. 
Yet one has to be aware that the orbital-selective(-like) scenario allied with 
intriguing fermiology in the intermediate doping region is quite specific, and to 
our knowledge has not yet been reported for other nickelates. 
This brings us surely to the most pressing
question concerning the pairing mechanism that emerges from the present 
normal-state scenario. A straightforward simple answer is not available, due to the
complex entanglement between localized and itinerant mutli-orbital degress of
freedom. From the renormalized-band perspective, a nesting mechanism involving
the flattened larger hole-pocket sheets around A might be conceivable. From a 
more localized perspective, recent suggestions based on Ni-$e_g$ couplings have been put
forward~\cite{zhav20,wer20}. However, those interesting proposals miss out on the
details of the intricate low-energy dispersions that evidently mark the intermediate 
doping region.

Beyond speculations, as the main conclusion, the superconductivity in thin films 
of Nd$_{1-x}$Sr$_x$NiO$_2$ is most definitely not of single-orbital cuprate kind. 
This should not be seen as a disappointment, but quite on the contrary, appreciated 
as the opening of a new fascinating chapter in the research on superconducting correlated
materials.

\begin{acknowledgments}
The author thanks K. Held, H. Y. Hwang, A. I. Lichtenstein, L. de' Medici,
A. J. Millis and P. Werner for helpful discussions. 
Financial support from the German Science Foundation (DFG) via the project 
LE-2446/4-1 is gratefully acknowledged. This work benefited from discussions 
held at the Aspen Center for Physics, which is supported by National Science 
Foundation grant PHY-1607611.
Computations were performed at the University of Hamburg and the JUWELS 
Cluster of the J\"ulich Supercomputing Centre (JSC) under project number hhh08.
\end{acknowledgments}

\bibliographystyle{new}
\bibliography{biblow}

\end{document}